# Nanoscale magnetic resonance imaging of proteins in a single cell

Pengfei Wang[1,2,3]*, Sanyou Chen[1,2,3]*, Maosen Guo[1,2,3]*, Shijie Peng[1,2,3]*, Mengqi Wang[1,2,3], Ming Chen[1,2,3], Wenchao Ma[1,2,3], Rui Zhang[4], Jihu Su[1,2,3], Xing Rong[1,2,3], Fazhan Shi[1,2,3], Tao Xu[4,5]†, Jiangfeng Du[1,2,3]†

[1]CAS Key Laboratory of Microscale Magnetic Resonance and Department of Modern Physics, University of Science and Technology of China, Hefei 230026, China.

[2]Hefei National Laboratory for Physical Sciences at the Microscale, University of Science and Technology of China, Hefei 230026, China.

[3]Synergetic Innovation Center of Quantum Information and Quantum Physics, University of Science and Technology of China, Hefei 230026, China.

[4]National Laboratory of Biomacromolecules, Institute of Biophysics, Chinese Academy of Sciences, Beijing 100101, China.

[5]College of Life Sciences, Chinese Academy of Sciences, Beijing 100049, China.

*These authors contributed equally to this work.

†Correspondence: djf@ustc.edu.cn, xutao@ibp.ac.cn.

**Magnetic resonance imaging (MRI) is a non-invasive and label-free technique widely used in medical diagnosis and life science research, and its success has benefited greatly from continuing efforts on enhancing contrast and resolution. Here we reported nanoscale MRI in a single cell using an atomic-size quantum sensor. With nitrogen-vacancy center in diamond, the intracellular protein ferritin has been imaged with a spatial resolution of ~ 10 nanometers, and ferritin-containing organelles were co-localized by correlative MRI and electron microscopy. Comparing to the current micrometer resolution in current state-of-art conventional MRI, our approach represents a 100-fold enhancement, and paves the way for MRI of intracellular proteins.**

Efforts to improve MRI spatial resolution has been essential for ensuring its continuing success. One approach for resolution improvement is to increase the higher external magnetic field. However, even at an external field of 14.1 Tesla, resolution of current MRI is at ~ 1 micrometer (*1*) as limited by electrical detection scheme (*2*). To break the resolution barrier, a series of MRI techniques, including super conducting quantum inference device (*3*) and magnetic resonance force microscopy (*4*), have been developed to detect and image the electron and nuclear spins at nanoscale resolution. However, the reported schemes require cryogenic environment and high vacuum, which severely curtail their applicability for imaging biological systems.

A recently developed quantum sensing method based on the nitrogen-vacancy (NV) center in diamond (*5, 6*) has radically pushed the boundary of magnetic resonance techniques. NV detections at the nanoscale and single-molecule level have been reported for organic samples (*7-9*) as well as proteins (*10, 11*) *in vitro*. By combining NV centers with scanning probe microscopy,

MRI with nanometer resolution have been demonstrated for single electron spin (*12*), small nuclear spin ensemble (*13, 14*), and paramagnetic metal ions (*15*). The NV center has also been used as a biocompatible magnetometer to image cells in a non-invasive manner at the subcellular scale, i.e., 0.4 μm (*16*). However, nanoscale MRI of single cells has not yet been reported, and the primary remaining barrier is accessibility of the samples to a scanning NV center.

We reported here two technical advancements to enable nanoscale MRI of intracellular proteins in a cell. In the studies, the cell was fixed to solid state, then sectioned to a cube shape and placed at a tuning fork scanning probe of an atomic force microscope (AFM), with the flat cross section of the cell exposed to the air (Fig. 1A). This sample placement set-up allowed positioning of the NV sensor within 10 nanometers to the target proteins, and the use of the AFM to suppress the thermal drift in sample positioning. In addition, by fabricating trapezoidal-cylinder shaped nanopillars at a bulk diamond surface and implanting NV centers at the top of the pillars (Fig. 1A), acquisition time for a MRI image was shortened by nearly one order (*17, 18*). Using this experimental setup, *in-situ* imaging of the intracellular protein ferritin was carried out.

Ferritin is a globular protein complex with an outer diameter of 12 nm (Fig. 1B), and within its 8-nm-diameter cavity the protein can house up to 4,500 iron atoms (*19*). The magnetic noise of the ferric ions can be detected by their effects on the $T_1$ relaxation time of a NV center. Previous work has clearly shown that the presence of ferritin decreases $T_1$ of the NV centers (*20-22*), and we confirmed this observation by fluorescence measurements of the time-dependence decay of the population of the $m_S = 0$ state of NV centers on a diamond surface coated with ferritins (Fig. 1C). Furthermore, it can be detected in a label-free manner using NV center (*20-22*) and transmission electron microscopy (TEM). This enabled the development of a correlated NV-MRI and TEM scheme to obtain and verify the first nanoscale MRI of a protein *in-situ*.

This work used hepatic carcinoma HepG2, a widely used cell line for the iron metabolism study. The cells were treated with ferric ammonium citrate (FAC) (Fig. 2A). As a result, the amount of ferritin increased significantly as verified by the results of confocal microscopy (CFM), western blot and TEM (Fig. S3) (*23*). Fig. 2B and Fig. S3E showed that ferritins localized primarily in intracellular puncta around the nucleus among the cytoplasm. Bulk electron paramagnetic resonance (EPR) spectroscopy confirmed the paramagnetic properties of ferritins in FAC-treated HepG2 cells (Fig. 2C), and mass spectrometry measurements precluded interference due to other paramagnetic metal ions (*23*).

All intracellular components of the Fe-loaded cells were immobilized by ultrafast high-pressure freezing. This treatment stabilized the intracellular structures and molecules and minimized Brownian motions that cause the proteins to randomly move on the scale up to a hundred nanometers (*24*). The frozen cells were embedded and polymerized in the LR White medium with the concerned intracellular structures remaining *in situ*. The embedded cell sample was then glued on the AFM tuning fork and shaped to the frustum of a prism with a few cells on the tip. The tip surface was sectioned to nanometer flatness by a diamond knife on the ultramicrotome and examined under AFM (*23*). During this sectioning step, a cell was dissected with a certain amount of ferritins remained just on or a few nanometers below the surface. The tuning fork was then placed on a sample stage, which contained a two-dimensional tilt that allows fine adjustments of the relative angle between the cell and diamond surface. After the adjustment, the nanopillar can parallel contact the surface of the cell cube.

NV-MRI images of ferritins were acquired by scanning the cell cube cross the diamond nanopillar while simultaneously measuring the NV spin depolarization rate under "leapfrog" scanning mode. Given the hardness of the diamond, the "leapfrog" mode maximally reduced the abrasion of the cell samples and allowed the reproduction of the results (*23*). Instead of measuring the whole depolarization curve, at each pixel, fluorescence decay was measured at a fixed free evolution time of $\tau = 50$ μs to reveal the degree of NV sensor spin polarization, which is correlated with the amount of ferritin within the sensing volume. The total time for one whole scan to acquire an image was about 2 hours, during which the positioning thermal drift was only few tens of nanometers.

A correlated NV-MRI and TEM scheme was developed to speed up and verify the nanoscale MRI images. We first cut off a cell section with a thickness of about 100 nm, and pre-locate the ferritin clusters by TEM (Fig. 3A). Then with NV-MRI setup, we relocated these clusters and scanned near their locations to search for the spin signal. Figure.3 shows an example of NV-MRI and TEM correlated result of intracellular ferritins with a pixel size of 43 nm. The diameter of ferritin clusters was between 100 and 500 nm. Given the different axial detection ranges between TEM (here, ~100 nm) and NV-MRI (typically 10 nm), the respective images showed micro-domains with different distributions of ferritins in a membrane-bound organelle. Some of the clusters appeared in both TEM and MRI images, while others were not observed in NV-MRI (Fig. 3B–D). Control measurements from CFM, EPR, TEM and mass spectrometry confirmed that the spin noise causing the depolarization of NV center was indeed from the intracellular ferritins.

To obtain the detail of ferritins in clusters, we acquired a high-resolution MRI image by minimizing the pixel size to 8.3 nm (Fig. 4). The image shows ferritins in a cluster of ~100 nm with a signal-to-noise (peak to peak) ratio >2. We observed several transitions from the background to the ferritin cluster with a minimal rising edge width of 8.3 nm, which was the same size as a single pixel (Fig. 4C). The spin noise from the intracellular ferritins was measured at around 0.02 $mT^2$. Using a model of 4,000 $Fe^{3+}$ uniformly distributed inside an 8 nm ferritin spherical core, the measured magnetic noise gives an estimation of approximately 9 nm for the vertical distance between the NV sensor and the center of the shallowest ferritin. We further calculated the point spread function of the measured signal and found that the spatial resolution determined by full width at half maximum (FWHM) was ~10 nm. This was close to the size of a single ferritin core, and agrees nicely with the observed sharp transition at the edge of the pixel (Fig. 4C). Consistent with these analyses, other lines did not show this sudden transition (*23*). This is mostly because the ferritins on the cluster boundary were less shallow beneath the cell section surface, and the signal dwindled as the ferritin was farther from the NV sensor.

NV centers have been shown to be a sensitive and appropriate sensor for the applications in biology. Here we developed NV-based methodologies to obtain the first 10 nm-resolution MRI image of a protein *in situ*. This method will contribute to further studies on the iron storage and release in cells, and the regulatory mechanism of iron metabolism disorders relating to iron at nanoscale, such as hemochromatosis, anemia, liver cirrhosis and Alzheimer's disease (*25, 26*). With additional improvements in stability and sensitivity, we can speed up the scanning, and image larger region of interest and obtain more information in the cell. Further studies will explore additional suitable target for correlated NV-MRI and TEM or optical microscopy detection, measurements of nuclear spin signal (*7-9*), as well as three dimensional cell tomography (*27*). The work reported here represents a significant step forward in pushing the boundary of MRI on endogenous intracellular nanostructure and macromolecules.

**Acknowledgments:** We thank Prof. Chongwen Zou for his help on the diamond annealing, Prof. Peter Zhifeng Qin for his help on language improvement of the manuscript, Prof. Li Bai and Dr. Xiaolan Xu for their help on cell culture and western blot. The fabrication of diamond nanopillars was performed at the USTC Center for Micro and Nanoscale Research and Fabrication.

**Fundation:** The authors in USTC are supported by the National Key R&D Program of China (Grant No. 2018YFA0306600 and 2016YFA0502400), the National Natural Science Foundation of China (Grants No. 81788101, 11722544, 11227901, 31470835, 91636217, 11761131011, and 31600685), the CAS (Grants No. GJJSTD20170001 and QYZDY-SSW-SLH004), the Anhui Initiative in Quantum Information Technologies (Grant No. AHY050000), the Anhui Provincial Natural Science Foundation (Grants No. 1808085J09 and 1608085QC63) and the Fundamental Research Funds for the Central Universities. The authors in Institute of Biophysics are supported by the Ministry of Science and Technology Program (Grants No. 2016YFA0500203), and by the National Natural Science Foundation of China (Grants No. 31421002 and 31730054).

**Author Contributions** J.D. and T.X. supervised the project and proposed the idea. J.D. and T.X. designed the proposal. P.W., M.G., S.P. and F.S. built the experiment set-up and carried out the NV-MRI experiments. S.C., R.Z. and M.G. prepared and characterized the cell samples. S.C., M.C., J.S. and X.R. measured the EPR and mass spectra. M.W. fabricated and imaged the diamond sensors. P.W., M.G. and W.M. simulated the data. P.W., S.C., M.G., F.S., T.X., and J.D. wrote the 6manuscript. All of authors participated in discussion and analysis over the data and manuscript.

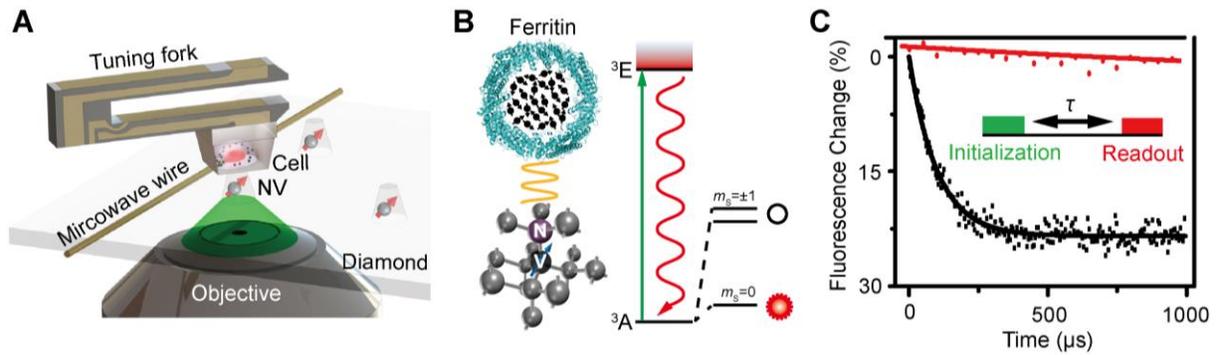

**Fig. 1. Schematic of the setup and experimental principle.** (**A**) Schematic view of the experimental setup. The cell embedded in resin is attached to a tuning fork and scans above the diamond nanopillar that contains a shallow NV center. A copper wire is used to deliver the microwave pulse to the NV center. A green laser (532 nm) from the CFM is used to address, initialize and readout the NV center. (**B**) Left: crystal lattice and energy level of the NV center. The NV center is a point defect that consists of a substitutional nitrogen atom and an adjacent vacancy in diamond. Right: schematic view of a ferritin. The black arrows indicate the electron spins of $Fe^{3+}$. (**C**) Experimental demonstration of the spin noise detection with and without ferritin in the form of polarization decay for the same NV center. The inset is the pulse sequence for detection and imaging of the ferritin. A 5 µs green laser is used to initialize the spin state to $m_s = 0$, followed by a free evolution time τ to accumulate the magnetic noise, and finally the spin state is read out by detecting the fluorescence intensity. The pulse sequence is repeated about $10^5$ times to acquire a good signal to noise ratio. The relaxation time is fitted to be 0.1 ms and 3.3 ms by exponential decay, respectively, indicating a spin noise of 0.01 $mT^2$.

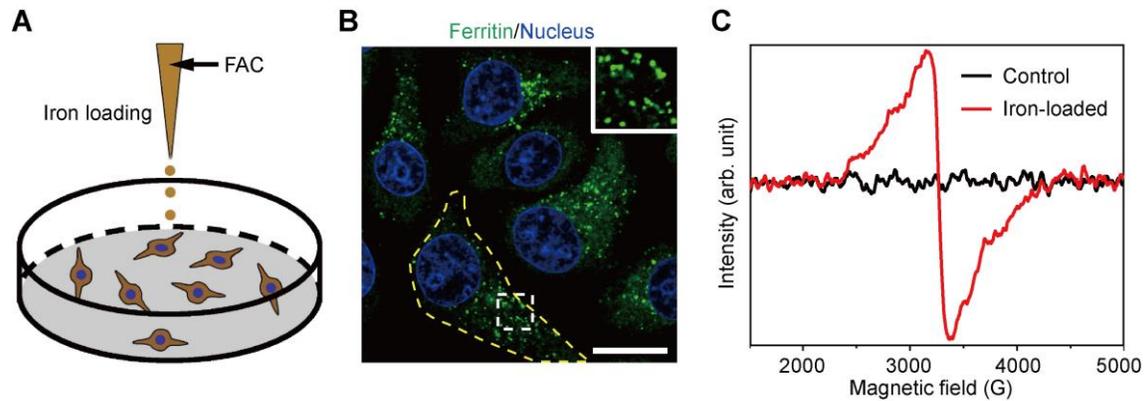

**Fig. 2. The preparation and characterization of ferritin-rich HepG2 cell samples.** (**A**) Schematic view of the treatment to cultured cells. Following the iron-loading or no treatment, the HepG2 cells were examined for fluorescence images and EPR spectra, respectively. For the MRI and TEM imaging, cell samples were treated through high-pressure freezing, freeze substitution, and sectioning. (**B**) A representative CFM image of ferritin structures (green) in iron-loaded HepG2 cells. The ferritin proteins were immunostained by anti-ferritin light chain (FTL) antibody. The nuclei are indicated by DAPI in the blue channel. Inset displays magnified ferritin structures. The yellow dashed line outlines the contour of a cell. Scale bar, 20 μm. (**C**) The electron paramagnetic resonance (EPR) spectra of control and iron-loaded HepG2 cells at $T = 300$ K. The HepG2 cells, harvested by trypsinization, were fixed with paraformaldehyde and immersed in Phosphate Buffered Saline (PBS).

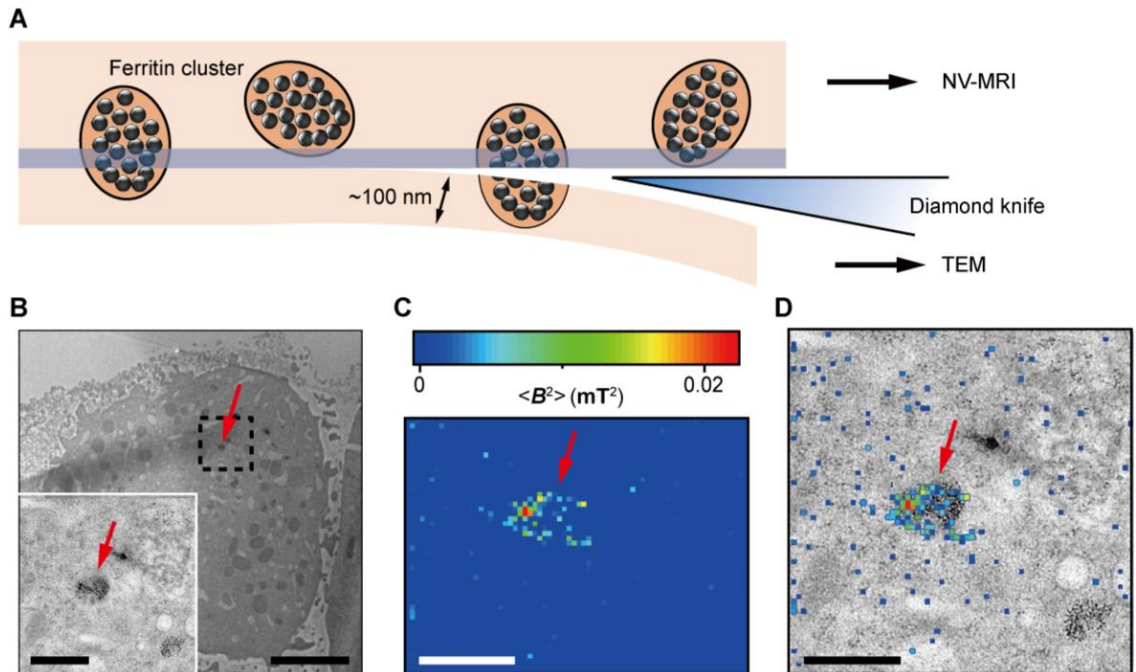

**Fig. 3. Correlative NV-MRI and TEM images.** (**A**) Schematic view of sectioning for correlative MRI and TEM imaging. The last section and the remained cube were transferred for TEM imaging and MRI scanning, respectively. The sectioning resulted in some split ferritin clusters that could be imaged under both microscope. The transparent blue strip of ~10 nm indicates the imaging depth of the NV-MRI, while in the TEM, the imaging depth is ~100 nm. (**B**) The distribution of ferritins from the last ultrathin section under TEM. Inset: The magnified figure of the part in black dashed box. (**C**) MRI result of the remained cell cube. The pixel size is 43 nm. (**D**) The merged NV-MRI and TEM micrograph shows ferritins in a membrane-bound organelle. The red arrows in B, C and D indicate the same ferritin cluster. Scale bars, 5 μm (B); 1 μm (Inset in (B), C and D).

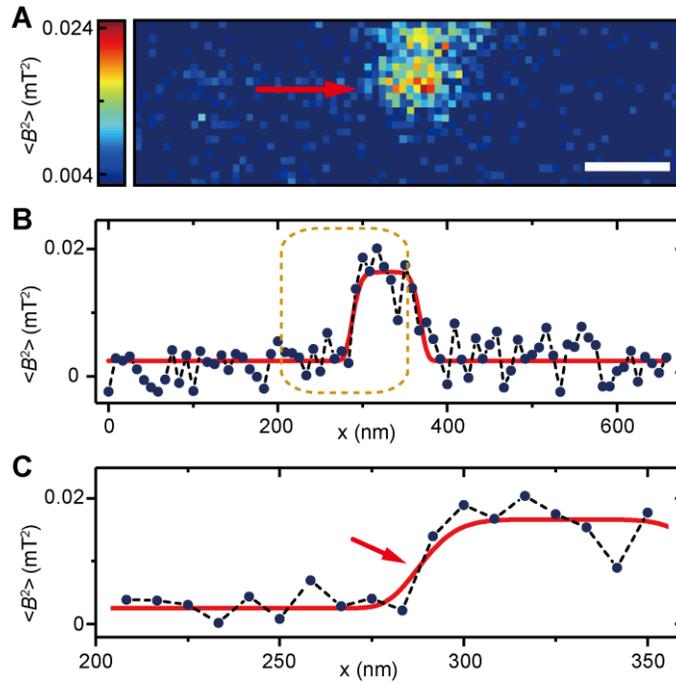

**Fig. 4. High-resolution MRI image of an intracellular ferritin cluster.** (**A**) A ferritin cluster imaged by the NV sensor with 80 ×24 pixels and pixel size of 8.3 nm. Scale bar, 100 nm. (**B**) The trace data of the scanning line in (A) directed by the red arrow. The platform indicates the ferritin cluster. The red curve fitted by a plateau function shows a guide to the eye. (**C**) The magnified figure of the gold dashed box in (B). The sharp transition indicated by the red arrow around x = 283 nm shows the scanning from the blank area to the area with ferritins.